\documentclass[preprints,article,accept,moreauthors,pdftex]{mdpi}

\firstpage{1} 
\pubvolume{xx}
\issuenum{1}
\articlenumber{5}
\pubyear{2019}
\copyrightyear{2019}
\history{Received: date; Accepted: date; Published: date}
\pdfoutput=1

\newcommand{\thh}{\textsuperscript{th}}
\newcommand{\pbar}{\bar{\mathbf p}}
\newcommand{\fbar}{\bar{f}}

\newcommand{\diff}{\;\mathrm{d}}
\newcommand{\Det}[1]{\left|#1\right|}

\Title{Multiplicative Decomposition of Heterogeneity in Mixtures of Continuous Distributions}

\Author{
Abraham Nunes$^{1,\dagger,*}$\orcidA{}, 
Martin Alda\textsuperscript{1}\orcidI{}
and Thomas Trappenberg\textsuperscript{2}\orcidH{}}

\AuthorNames{Abraham Nunes, Martin Alda, and Thomas Trappenberg}

\address{$^{1}$ Department of Psychiatry, Dalhousie University, Halifax, Nova Scotia, Canada\\
$^{2}$ Faculty of Computer Science, Dalhousie University, Halifax, Nova Scotia, Canada}

\corres{Correspondence: \href{mailto:nunes@dal.ca}{nunes@dal.ca}}

\firstnote{Current address: 5909 Veterans Memorial Lane (8th Floor), Abbie J. Lane Memorial Building, QE I.I. Health Sciences Centre, Halifax, Nova Scotia, B3H 2E2, Canada} 

\abstract{A system's heterogeneity (\textit{diversity}) is the effective size of its event space, and can be quantified using the R\'enyi family of indices (also known as Hill numbers in ecology or Hannah-Kay indices in economics), which are indexed by an elasticity parameter $q \geq 0$. Under these indices, the heterogeneity of a composite system (the $\gamma$-heterogeneity) is decomposable into heterogeneity arising from variation \textit{within} and \textit{between} component subsystems (the $\alpha$- and $\beta$-heterogeneity, respectively). Since the average heterogeneity of a component subsystem should not be greater than that of the pooled system, we require that $\gamma \geq \alpha$. There exists a multiplicative decomposition for R\'enyi heterogeneity of composite systems with discrete event spaces, but less attention has been paid to decomposition in the continuous setting.  We therefore describe multiplicative decomposition of the R\'enyi heterogeneity for continuous mixture distributions under parametric and non-parametric pooling assumptions. Under non-parametric pooling, the $\gamma$-heterogeneity must often be estimated numerically, but the multiplicative decomposition holds such that $\gamma \geq \alpha$ for $q > 0$. Conversely, under parametric pooling, $\gamma$-heterogeneity can be computed efficiently in closed-form, but the $\gamma \geq \alpha$ condition holds reliably only at $q=1$. Our findings will further contribute to heterogeneity measurement in continuous systems.}


\usepackage{todonotes}

\begin{document}
%%%%%%%%%%%%%%%%%%%%%%%%%%%%%%%%%%%%%%%%%%

%%%%%%%%%%%%%%%%%%%%%%%%%%%%%%%%%%%%%%%%%%

\section{Introduction}\label{s:introduction}

Measurement of heterogeneity is important across many scientific disciplines. Ecologists are interested in the heterogeneity of ecosystems' biological composition (biodiversity) \cite{Hooper2005}, economists are interested in the heterogeneity of resource ownership (wealth equality) \cite{Cowell2011}, and medical researchers and physicians are interested in the heterogeneity of diseases and their presentations \cite{Nunes2020}. Using R\'enyi heterogeneity \cite{Nunes2020, Nunes2020_DMH, Nunes2020_RRH}, which for categorical random variables corresponds to ecologists' \textit{Hill numbers} \cite{Hill1973} and economists' \textit{Hannah-Kay indices} \cite{Hannah1977}, one can measure a system's heterogeneity as its effective number of distinct configurations.

%\citet{Nunes2020_RRH} showed that heterogeneity measurement in various scientific disciplines can be generalized by \textit{representational R\'enyi heterogeneity} (RRH). In RRH, the heterogeneity of a system (random variable) $X$ with observable event space $\mathcal X$ is measured as the effective size of $\mathcal X$ with respect to a latent or unobservable \textit{representation} $\mathcal Z$ that is either pre-specified or learned. For example, in an ecological survey, the space of observable organisms (whether images, live samples, or otherwise) would be denoted $\mathcal X$. Let the representation be a categorical space of species labels $\mathcal Z = \left\{1,2,\ldots,S\right\}$, over which there is a family of probability distributions, $\mathcal P(\mathcal Z)$. The probabilistic representation of an observed organism is then obtained via vector-valued function $\mathbf f:\mathcal X \to \mathcal P(\mathcal Z)$. However, if organism labeling is deterministic---that is, where $\mathcal P(\mathcal Z)$ contains only the degenerate distribution---then RRH recovers the traditional Hill numbers \cite{Hill1973, Nunes2020_RRH}. Conversely, if $\mathcal X$ is the space of equal-valued economic assets, $\mathcal Z$ is a categorical space identifying asset owners (agents), and $\mathbf f:\mathcal X \to \mathcal P(\mathcal Z)$ is a function mapping an observed asset $x \in X$ onto a degenerate distribution over agents, then RRH recovers the Hannah-Kay indices \cite{Hannah1977, Nunes2020_RRH}. 

The heterogeneity of a mixture or ensemble of systems is often known as $\gamma$-heterogeneity, and is generated by variation occurring \textit{within} and \textit{between} constituent subsystems. A good heterogeneity measure will facilitate decomposition of $\gamma$-heterogeneity into $\alpha$ (within subsystem) and $\beta$ (between subsystem) components. Under this decomposition, we require that $\gamma \geq \alpha$, since it is counterintuitive that the heterogeneity of the overall ensemble should be less than any of its constituents, let alone the ``average'' subsystem \cite{Lande1996, Jost2007}. Such a decomposition was introduced by \citet{Jost2007} for systems represented on discrete event spaces (such as representations of organisms by species labels). However, many data are better modeled by continuous embeddings; including word semantics \cite{Mikolov2013a, Pennington2014, Nickel2017}, genetic population structure \cite{Price2006}, and natural images \cite{Karras2019stylegan2}. Unfortunately, there is considerably less understood about how to decompose R\'enyi heterogeneity in such cases where data are represented on non-categorical spaces \cite{Nunes2020_DMH}. Although there are decomposable functional diversity indices expressed in numbers equivalent, they require categorical partitioning of the data (in order to supply species (dis)similarity matrices) \cite{Ricotta2009, Leinster2012a, Chiu2014a, Chao2019} and setting sensitivity or threshold parameters for (dis)similarities \cite{Leinster2012a, Chao2019}. For many research applications, such as those in psychiatry \cite{Nunes2020, Nunes2020_DMH, Marquand2016} or involving unsupervised learning \cite{Price2006, Karras2019stylegan2}, we may not have categorical partitions of the observable space that are valid, reliable, and of semantic relevance. If we are to apply R\'enyi heterogeneity to such continuous-space systems, then we must demonstrate that its multiplicative decomposition of $\gamma$-heterogeneity into $\alpha$ and $\beta$ components is retained. 

Therefore, our present work extends the \citet{Jost2007} multiplicative decomposition of R\'enyi heterogeneity to the analysis of continuous systems, and provides conditions under which the $\gamma\geq\alpha$ condition is satisfied. In Section \ref{s:background}, we introduce decomposition of the R\'enyi heterogeneity in categorical and continuous systems. Specifically, we highlight that the most important decision guiding the availability of a decomposition is how one defines the distribution over the mixture of subsystems. We show that for non-parametrically pooled systems (i.e. finite mixture models, illustrated in Section \ref{s:gmm}), the $\gamma \geq \alpha$ condition can hold for all values of the R\'enyi elasticity parameter $q > 0$, but that $\gamma$-heterogeneity will generally require numerical estimation. Section \ref{s:hier-gaussian-random-effects} introduces decomposition of R\'enyi heterogeneity under parametric assumptions on the pooled system's distribution. In this case, which amounts to a Gaussian mixed-effects model (as commonly implemented in biomedical meta-analyses), we show that $\gamma\geq\alpha$ will hold at $q=1$, though not necessarily at $q\neq1$. Finally, in Section \ref{s:discussion}, we discuss the implications of our findings and scenarios in which parametric or non-parametric pooling assumptions might be particularly useful.

%%%%%%%%%%%%%%%%%%%%%%%%%%%%%%%%%%%%%%%%%%

\section{Background}
\label{s:background}

\subsection{Categorical R\'enyi Heterogeneity Decomposition}
\label{ss:categorical-renyi-decomposition}

In this section, we consider the definition and decomposition of R\'enyi heterogeneity for a composite random variable (or ``system'') that we call a \textit{discrete mixture} (Definition \ref{def:discrete-mixture}).

\begin{Definition}[Discrete Mixture]\label{def:discrete-mixture}
A random variable or system $X$ is called a discrete mixture when it is defined on an $n$-dimensional discrete state space $\mathcal X = \{1,2,\ldots,n\}$ with probability distribution $\pbar = \left(\bar{p}_{i}\right)_{i=1,2,\ldots,n}$, where $\bar{p}_i$ is the probability that $X$ is observed in state $i \in \mathcal X$. Furthermore, let $X$ be an aggregation of $N$ component subsystems $X_1, X_2, \ldots, X_N$ with corresponding probability distributions $\mathbf P = \left(p_{ij}\right)_{i=1,2,\ldots,N}^{j=1,2,\ldots,n}$. The proportion of $X$ attributable to each component is governed by the weights $\mathbf w = \left(w_i\right)_{i=1,2,\ldots,N}$, where $0 \leq w_i \leq 1$ and $\sum_{i=1}^N w_i = 1$.
\end{Definition}

Let $X$ be a discrete mixture. The R\'enyi heterogeneity for the $i$\thh component is 

\begin{equation}
    \Pi_q\left(X_i\right) = \left(\sum_{j=1}^n p_{ij}^q \right)^{\frac{1}{1-q}},
    \label{eq:renyi-het}
\end{equation}

\noindent which is the effective number of states in $X_i$. 
Assuming the pooled distribution over discrete mixture $X$ is a weighted average of subsystem distributions, $\pbar = \mathbf P^\top \mathbf w$, the $\gamma$-heterogeneity is thus

\begin{equation}
    \Pi_q^\gamma \left(X\right) = \left(\sum_{i=1}^n \bar{p}_i^q \right)^{\frac{1}{1-q}},
    \label{eq:gamma-het}
\end{equation}

\noindent which we interpret as the effective number of states in the pooled system $X$. 

\citet{Jost2007} proposed the following decomposition of $\gamma$-heterogeneity:

\begin{equation}
    \Pi_q^{\gamma}\left(X\right) = \Pi_q^\alpha \left(X\right) \Pi_q^{\beta} \left(X\right),
    \label{eq:jost-decomposition}
\end{equation}

\noindent where $\Pi_q^\alpha\left(X\right)$ and $\Pi_q^\beta \left(X\right)$ are summary measures of heterogeneity due to variation \textit{within} and \textit{between} subsystems, respectively. Since the $\gamma$ factor has units of effective number of states in the pooled system, and $\alpha$ has units of effective number of states \textit{per component}, then 

\begin{equation}
    \Pi_q^\beta\left(X\right) = \frac{\Pi_q^\gamma\left(X\right)}{\Pi_q^\alpha\left(X\right)}
    \label{eq:beta-het}
\end{equation}

\noindent yields the effective number of components in $X$. 

For discrete mixtures, \citet{Jost2007} specified the functional form for $\alpha$-heterogeneity as 

\begin{equation}
    \Pi_q^\alpha\left(X\right) = \left\{
    \begin{array}{ll} 
    \left(\sum_{i=1}^N \frac{ w_i^q}{\sum_{k=1}^N w_k^q} \sum_{j=1}^n p_{ij}^q\right)^{\frac{1}{1-q}} & q\neq 1 \\
    \exp\{- \sum_{i=1}^N w_i \sum_{j=1}^n p_{ij}\log p_{ij}\} & q = 1
    \end{array}\right.,
    \label{eq:cat-renyi-alpha}
\end{equation}

\noindent which allows the decomposition in Equation \ref{eq:jost-decomposition} to satisfy the following desiderata: 

\begin{enumerate}
    \item The $\alpha$ and $\beta$ components are independent \cite{Wilson1984}
    \item The within-group heterogeneity is a lower bound on total heterogeneity \cite{Lande1996}: $\Pi_q^\alpha \leq \Pi_q^\gamma$
    \item The $\alpha$-heterogeneity is a form of average heterogeneity over groups
    \item The $\alpha$ and $\beta$ components are both expressed in numbers equivalent.
\end{enumerate}

\noindent Specifically, \citet{Jost2007} proved that $\Pi_q^\gamma\left(X\right) \geq \Pi_q^\alpha\left(X\right)$ is guaranteed for all $q \geq 0$ when $w_i = w_j$ for all $(i,j) \in \{1,2,\ldots,N\}$, or for unequal weights $\mathbf w$ if the elasticity is set to the Shannon limit of $q\to1$.

\subsection{Continuous R\'enyi Heterogeneity Decomposition}
\label{ss:continuous-renyi-decomposition}

Let $X$ be a non-parametric continuous mixture according to Definition \ref{def:nonparametric-continuous-mixture}. Despite individual mixture components in $X$ potentially having parametric probability density functions, we call this a ``non-parametric'' mixture because the distribution over \textit{pooled} components does not assume the form of a known parametric family. 

\begin{Definition}[Non-parametric Continuous Mixture] \label{def:nonparametric-continuous-mixture}
A non-parametric continuous mixture is a random variable $X$ defined on an $n$-dimensional continuous space $\mathcal X \subseteq \mathbb R^n$, and composed of subsystems $X_1, X_2, \ldots, X_N$, with respective probability density functions $\mathbf f(\mathbf x) = \left\{f_i(\mathbf x)\right\}_{i=1,2,\ldots,N}$ and weights $\mathbf w = \left(w_i\right)_{i=1,2,\ldots,N}$ such that $\sum_{i=1}^N w_i = 1$ and $0 \leq w_i \leq 1$. The pooled probability density over $X$ is defined as 

\begin{equation}
    \fbar(\mathbf x) = \sum_{i=1}^N w_i f_i(\mathbf x).
    \label{eq:pooled-density}
\end{equation}

\end{Definition}

The continuous R\'enyi heterogeneity for the $i$\thh subsystem of $X$ is

\begin{equation}
    \Pi_q\left(X_i\right) = \left(\int_{\mathcal X} f_i^q(\mathbf x) \diff \mathbf x \right)^{\frac{1}{1-q}},
    \label{eq:single-continuous-renyi}
\end{equation}

\noindent whose interpretation is given by Proposition \ref{prop:continuous-renyi} (see Proposition A3 in \citet{Nunes2020_RRH} for the proof), which we henceforth call the ``effective volume'' of the event space or domain of $X_i$.

\begin{Proposition}[R\'enyi Heterogeneity of a Continuous Random Variable] \label{prop:continuous-renyi} The R\'enyi heterogeneity of a continuous random variable $X$ defined on event space $\mathcal X \subseteq \mathbb R^n$ with probability density function $f$ is equal to the magnitude of the volume of an $n$-cube over which there is a uniform probability density with the same R\'enyi heterogeneity as that in $X$.
\end{Proposition}

Given the pooled distribution as defined in Equation \ref{eq:pooled-density}, the R\'enyi heterogeneity over the mixture, which is the $\gamma$-heterogeneity, is

\begin{equation}
    \Pi_q^\gamma\left(X\right) = \left(\int_{\mathcal X} \fbar^q(\mathbf x) \diff \mathbf x\right)^{\frac{1}{1-q}}.
    \label{eq:pooled-continuous-renyi}
\end{equation}

\noindent The $\gamma$-heterogeneity is thus the total effective volume of $X$'s domain. The $\alpha$-heterogeneity represents the effective volume per component mixture component in $X$, and is computed as follows:

\begin{equation}
    \Pi_q^\alpha\left( X \right) = \left( \sum_{i=1}^{N} \frac{w_i^q}{\sum_{k=1}^{N} w_k^q} \int_{\mathcal X} f_i^q(\mathbf x) \diff \mathbf x \right)^\frac{1}{1-q}.
    \label{eq:continuous-alpha-renyi}
\end{equation}

Given Equations \ref{eq:pooled-continuous-renyi} and \ref{eq:continuous-alpha-renyi}, the following theorem provides conditions under which $\gamma \geq \alpha$ is satisfied for a non-parametric continuous mixture. The proof is analogous to that given by \citet{Jost2007} for discrete mixtures, and is detailed in Appendix \ref{app:proofs}. 

\begin{Theorem} \label{thm:decomposition-nonparametric-continuous}
If $X$ is a non-parametric continuous mixture (Definition \ref{def:nonparametric-continuous-mixture}), with $\gamma$-heterogeneity specified by Equation \ref{eq:pooled-continuous-renyi} and $\alpha$-heterogeneity given by Equation \ref{eq:continuous-alpha-renyi}, then 

\begin{equation}
    \Pi_q^{\beta}\left(X\right) = \frac{\Pi_q^{\gamma}\left(X\right)}{\Pi_q^{\alpha}\left(X\right)} \geq 1
    \label{eq:thm-statement}
\end{equation}
    
\noindent under the following conditions: 

\begin{enumerate}
    \item $q = 1$
    \item $q > 0$ when weights are equal for all mixture components.
\end{enumerate}
\end{Theorem}

If $\int_{\mathcal X}f_i^q(x) \diff x$ is analytically tractable for all $i \in \{1,2,\ldots,N\}$, then a closed form expression for $\Pi_q^\alpha\left(X\right)$ will be available. If $\int_{\mathcal X} \fbar^q(x) \diff x$ is also analytically tractable, then so too will be $\Pi_q^\beta\left(X\right)$. However, this will depend entirely on the functional form of $\fbar$, and will rarely be the case using real world data. In the majority of cases, $\int_{\mathcal X} \fbar^q(x) \diff x$ will have to be computed numerically. 

%%%%%%%%%%%%%%%%%%%%%%%%%%%%%%%%%%%%%%%%%%

\section{R\'enyi Heterogeneity Decomposition under a Non-parametric Pooling Distribution}
\label{s:gmm}

Definition \ref{def:gaussian-mixture} defines a general Gaussian mixture $X$ as a weighted combination of component Gaussian random variables, without identifying the function form of the composition. The non-parametric Gaussian mixture, where the distribution over $X$ is a simple model average over it's Gaussian components, is specified in Definition \ref{def:nonparametric-gaussian-mixture}.

\begin{Definition}[Gaussian Mixture] \label{def:gaussian-mixture}
The $n$-dimensional Gaussian mixture $X$ is a weighted combination of the set of $n$-dimensional Gaussian random variables $\left\{X_i\right\}_{i=1,2,\ldots,N}$ with component weights $\mathbf w = \left(w_i\right)_{i=1,2,\ldots,N}$ such that $0 \leq w_i \leq 1$ and $\sum_{i=1}^N w_i = 1$. The probability density function of component $X_i$ is denoted $\mathcal N\left(\mathbf x|\boldsymbol\mu_i, \boldsymbol\Sigma_i\right)$, and is parameterized by an $n \times 1$ mean vector $\boldsymbol\mu_i$ and $n \times n$ covariance matrix $\boldsymbol\Sigma_i$.
\end{Definition}

\begin{Definition}[Non-parametric Gaussian Mixture] \label{def:nonparametric-gaussian-mixture}
We define the random variable $X$ as a non-parametric Gaussian mixture if it is a Gaussian mixture (Definition \ref{def:gaussian-mixture}) whose probability density function is defined as 

\begin{equation}
\fbar(\mathbf x|\boldsymbol\mu_{1:N}, \boldsymbol\Sigma_{1:N}, \mathbf w) = \sum_{i=1}^N w_i \mathcal N(\mathbf x|\boldsymbol\mu_i, \boldsymbol\Sigma_i),
\label{eq:gmm-likelihood}
\end{equation}

\noindent where $\boldsymbol\mu_{1:N}$ and $\boldsymbol\Sigma_{1:N}$ denote the set of component mean vectors $\boldsymbol\mu_{1},\ldots,\boldsymbol\mu_N$ and covariance matrices $\boldsymbol\Sigma_{1},\ldots,\boldsymbol\Sigma_N$, respectively.
\end{Definition}

We now introduce the R\'enyi heterogeneity of a single $n$-dimensional Gaussian random variable (Proposition \ref{prop:gaussian-renyi-het}) and subsequently characterize the $\gamma$-, $\alpha$-, and $\beta$-heterogeneity values for a non-parametric Gaussian mixture.

\begin{Proposition}[R\'enyi Heterogeneity of a Multivariate Gaussian]
\label{prop:gaussian-renyi-het}
The R\'enyi heterogeneity of an $n$-dimensional Gaussian random variable $X$ with mean $\boldsymbol\mu$ and covariance matrix $\boldsymbol\Sigma$ is

\begin{equation}
\Pi_q\left(X\right)=
\left\{
\begin{array}{ll}
\mathrm{Undefined} & q=0 \\
(2 \pi  e)^{\frac{n}{2}} \Det{\boldsymbol\Sigma}^\frac{1}{2} & q=1 \\
(2 \pi )^{\frac{n}{2}} \Det{\boldsymbol\Sigma}^\frac{1}{2} & q=\infty  \\
(2 \pi )^{\frac{n}{2}} q^{\frac{n}{2 (q-1)}} \Det{\boldsymbol\Sigma}^\frac{1}{2} & q\notin\{0,1,\infty\} \\
\end{array}\right..
\label{eq:cont-repamrenyi-gaussian}
\end{equation}
\end{Proposition}

The proof of Proposition \ref{prop:gaussian-renyi-het} is included in Appendix \ref{app:proofs}. Unfortunately, a closed form solution such as Equation \ref{eq:cont-repamrenyi-gaussian} cannot be obtained for the $\gamma$-heterogeneity of a non-parametric Gaussian mixture,

\begin{equation}
\Pi_q^{\gamma}\left(X\right) = \left( \int_{\mathcal X} \left( \sum_{i=1}^N w_i \mathcal N(\mathbf x|\boldsymbol\mu_i, \boldsymbol\Sigma_i) \right)^q \diff \mathbf x \right)^\frac{1}{1-q},
\label{eq:nonparam-gamma-het}
\end{equation}

\noindent which must be computed numerically to yield the effective size of the mixture's domain. This process may be computationally expensive, particularly in high dimensions. Conversely, Equation \ref{eq:continuous-alpha-renyi}, which yields the effective size of the domain per mixture component, can be evaluated in closed form for a Gaussian mixture:

\begin{equation}
\Pi_q^\alpha\left( X \right) = \left\{\begin{array}{ll}
\mathrm{Undefined} & q = 0 \\
\exp\left\{\frac{1}{2}\left(n +  \sum_{i=1}^N w_i \log \left|2 \pi \boldsymbol\Sigma_i\right|\right)\right\} & q = 1 \\
0 & q = \infty \\
(2\pi)^\frac{n}{2} \left(\sum_{i=1}^N \frac{w_i^q}{\sum_{j=1}^N w_j^q}
\frac{ \left|\boldsymbol\Sigma_i\right|^\frac{1}{2}}{q^\frac{n}{2}}\right)^\frac{1}{1-q} & q \notin \{0,1,\infty\}\\
\end{array}
\right..
\label{eq:alpha-heterogeneity-gaussian}
\end{equation}

The $\beta$-heterogeneity, which returns the effective number of components in the mixture, can then be computed using Equation \ref{eq:beta-het}. Example \ref{ex:gaussian-numerical} demonstrates an important property of considering $X$ as a \textit{non-parametric} Gaussian mixture: that low-probability regions of the domain \textit{between} well-separated components will have little to no effect on the $\gamma$- or $\beta$-heterogeneity estimates.

\begin{Example}[Decomposition of R\'enyi heterogeneity in a univariate Gaussian mixture]\label{ex:gaussian-numerical}
Consider three non-parametric Gaussian mixtures $X^{(1)}, X^{(2)}, X^{(3)}$ defined on $\mathbb R$ whose number of components are respectively $N_1 = 2$, $N_2 = 3$, and $N_3=4$. Components in each mixture are equally weighted---that is, the components of mixture $X^{(j)}$ have weights $w_i^{(j)} = 1/N_j$ for all $i \in \{1,2,\ldots,N_j\}$---and have equal standard deviation $\sigma = 0.5$. This yields a per-component R\'enyi heterogeneity of approximately 2.07, which is also consequently the $\alpha$-heterogeneity for each Gaussian mixture. 

Figure \ref{fig:gaussian-numerical-int} demonstrates the multiplicative decomposition of R\'enyi heterogeneity (at $q=1$) in these Gaussian mixtures, where $\gamma$-heterogeneity was computed numerically, across varying separations of respective mixtures' component means. Note that the $\beta$-heterogeneity in this case represents the effective number of distinct components in the mixture distribution, and is bound between 1 (when all components overlap), and $N_j$ (when all components are well separated). Further separating the mixture components beyond the point at which $\beta$-heterogeneity reaches $N_j$ yielded no further increase in $\beta$-heterogeneity.

\begin{figure}[H]
	\centering 
	\includegraphics[width=0.8\textwidth]{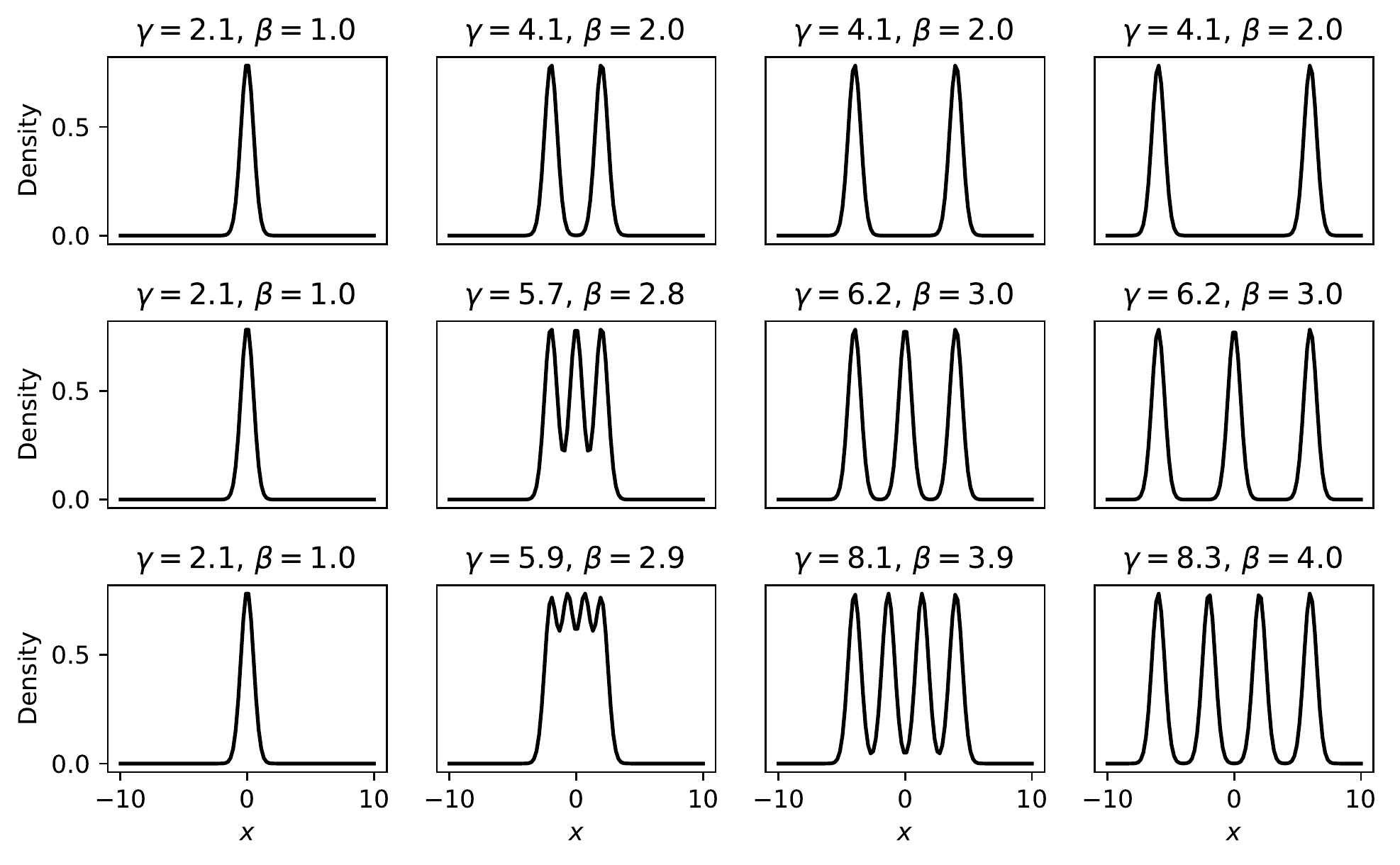}
	\caption{Demonstration of the multiplicative decomposition of R\'enyi heterogeneity in Gaussian mixture models, where $\gamma$-heterogeneity is computed using numerical integration. Each row represents a different number of mixture components (from top to bottom: 2, 3, and 4 univariate Gaussians with $\sigma=0.5$, respectively). Each column shows a case in which the component locations are progressively further separated ($\max_i \mu_i - \min_i \mu_i$ distance from left to right: 0, 2, 4, 6). The $\alpha$-heterogeneity in all scenarios was $\approx 2.07$. The headings on each panel show the resulting $\gamma$ and $\beta$-heterogeneity values.}
	\label{fig:gaussian-numerical-int}
\end{figure}

\end{Example}

Assuming sufficiently accurate approximation of the integral in Equation \ref{eq:nonparam-gamma-het}, the $\gamma$-heterogeneity in Example \ref{ex:gaussian-numerical} appears to reach a limit corresponding to the sum of effective domain sizes under all mixture components, and the $\beta$-heterogeneity reaches a limit corresponding to the number of individual mixture components.

Unfortunately, computation of $\beta$-heterogeneity in a non-parametric Gaussian mixture will yield results whose accuracy will depend on the error of numerical integration, and which may consume significant computational resources when evaluated for large $N$ (many components) and large $n$ (high dimension). Although the non-parametric pooling approach may be the only available method for many distribution classes, a computationally efficient parametric pooling approach exists for Gaussian mixtures, to which we now turn our attention.

\section{R\'enyi Heterogeneity Decomposition Under a Parametric Pooling Distribution}
\label{s:hier-gaussian-random-effects}

This section introduces the parametric Gaussian mixture (Definition  \ref{def:parametric-gaussian-mixture}), and subsequently provides conditions under which decomposition of its heterogeneity satisfies the requirement that $\alpha$-heterogeneity be a lower bound on $\gamma$-heterogeneity (Theorem \ref{thm:gmem-alphabound}).

\begin{Definition}[Parametric Gaussian Mixture]\label{def:parametric-gaussian-mixture} We define the random variable $X$ as an $n$-dimensional parametric Gaussian mixture if it is a Gaussian mixture (Definition \ref{def:gaussian-mixture}) whose probability density function is defined as 

\begin{equation}
\fbar(\mathbf x|\boldsymbol\mu_\ast, \boldsymbol\Sigma_\ast) = \mathcal N(\mathbf x|\boldsymbol\mu_\ast, \boldsymbol\Sigma_\ast),
\label{eq:gaussian-global}
\end{equation}

\noindent with pooled mean vector 

\begin{equation}
\boldsymbol\mu_\ast =  \sum_{i=1}^N w_i \boldsymbol\mu_i,
\label{eq:pooled-mean}
\end{equation}

\noindent and pooled covariance matrix

\begin{equation}
\boldsymbol\Sigma_\ast = - \boldsymbol\mu_\ast \boldsymbol\mu_\ast^\top + \sum_{i=1}^N w_i \left( \boldsymbol\Sigma_i  + \boldsymbol\mu_i \boldsymbol\mu_i^\top \right).
\label{eq:pooled-cov}
\end{equation}
\end{Definition}

The efficiency of assuming a parametric, rather than non-parametric, Gaussian mixture is that $\gamma$-heterogeneity for the latter may be computed in closed form using Equation \ref{eq:cont-repamrenyi-gaussian} (it is simply a function of Equation \ref{eq:pooled-cov}). However, the critical difference between the parametric and non-parametric Gaussian mixture assumptions is that $\gamma$-heterogeneity---and therefore $\beta$-heterogeneity---will depend on the component means $\boldsymbol\mu_{1:N}$, according to the following Lemma.

\begin{Lemma}[Relationship of $\gamma$-Heterogeneity to Component Dispersion]\label{lem:gamma-dispersion}
Let $X$ and $X^\prime$ be $N$-component parametric Gaussian mixtures on $\mathbb R^n$ with component-wise mean vectors $\boldsymbol\mu_{1:N} = \left\{\boldsymbol\mu_i\right\}_{i=1,2,\ldots,N}$ and $\boldsymbol\mu_{1:N}^\prime = \left\{\sqrt{c}\boldsymbol\mu_i\right\}_{i=1,2,\ldots,N}$, where $c \geq 1$ is a scaling factor. The component-wise weights $\mathbf w$ and covariance matrices $\boldsymbol\Sigma_{1:N} = \left\{ \boldsymbol\Sigma_i \right\}_{i=1,2,\ldots,N}$ are identical between $X$ and $X^\prime$. Finally, let $\boldsymbol\Sigma_\ast$ and $\boldsymbol\Sigma_\ast^\prime$ be the pooled covariance matrices for $X$ and $X^\prime$, respectively. Then, for all $c \geq 1$, we have that 

\begin{equation}
    \Pi_q^\gamma \left(X^\prime \right) \geq \Pi_q^\gamma \left(X\right),
    \label{eq:lem1-inequality}
\end{equation}

\noindent with equality if $c = 1$.

\end{Lemma}

Lemma \ref{lem:gamma-dispersion}, whose proof is detailed in Appendix \ref{app:proofs}, implies that the resulting $\beta$-heterogeneity of a parametric Gaussian mixture will increase as the mixture component means are spread further apart. This follows from the fact that Equation \ref{eq:alpha-heterogeneity-gaussian}, which is computed component-wise, remains a valid expression of the $\alpha$-heterogeneity in a parametric Gaussian mixture.

Before stating the conditions under which $\alpha$ is a lower bound on $\gamma$ for a parametric Gaussian mixture (Theorem \ref{thm:gmem-alphabound}), we introduce the following Lemma, whose proof is left to Appendix \ref{app:proofs}.

\begin{Lemma}\label{lem:power-mean-inequality}
If $\left\{\boldsymbol\Sigma_i\right\}_{i=1,2,\ldots,N}$ is a set of $N \in \mathbb N_{\geq 2}$ positive semidefinite $n \times n$ matrices with corresponding weights $\mathbf w = \left(w_i\right)_{i=1,2,\ldots,N}$ such that $0 \leq w_i \leq 1$ and $\sum_{i=1}^N w_i = 1$, then

\begin{equation}
    \Det{\sum_{i=1}^N w_i \boldsymbol\Sigma_i}^\frac{1}{2} \geq \sum_{i=1}^N \Det{\boldsymbol\Sigma_i}^\frac{w_i}{2}.
    \label{eq:power-mean-inequality}
\end{equation}
\end{Lemma}

\begin{Theorem}\label{thm:gmem-alphabound}
	The R\'enyi $\beta$-heterogeneity of order $q=1$ of a parametric Gaussian mixture $X$ (Definition \ref{def:parametric-gaussian-mixture}) has a lower bound of 1:
	
	\begin{equation}
	    \Pi_1^\beta\left(X\right) = \frac{\Pi_1^\gamma\left(X\right)}{\Pi_1^\alpha\left(X\right)} \geq 1
	    \label{eq:beta-ratio-thm1}
	\end{equation}
\end{Theorem}

\begin{proof}\label{proof:gmem-alphabound} 
    Recall that $\Pi_q^\alpha\left(X\right)$ is independent of the mean-vectors of components in $X$ (Equation \ref{eq:alpha-heterogeneity-gaussian}). Furthermore, it follows from Lemma \ref{lem:gamma-dispersion} that if $\boldsymbol\mu_{1:N} = \left\{\mathbf 0\right\}_{i=1,2,\ldots,N}$, where $\mathbf 0$ is an $n\times 1$ zero vector, then for any parametric Gaussian mixture $X^\prime$ with means $\boldsymbol\mu_{1:N}^\prime$ we will have $\Pi_q^\gamma\left(X^\prime\right) \geq \Pi_q^\gamma\left(X\right)$, where equality is obtained if $\boldsymbol\mu_{1:N}^\prime$ are also zero vectors, or the covariance of mean vectors in $X^\prime$,
    
    \begin{equation}
        \mathrm{Cov}[\boldsymbol\mu^\prime] = \mathbb E[\boldsymbol\mu^\prime {\boldsymbol\mu^\prime}^\top] - \mathbb E[\boldsymbol\mu^\prime] \mathbb E[\boldsymbol\mu^\prime]^\top,
    \end{equation}
    
    \noindent is otherwise singular. Thus, it suffices to prove our theorem under the assumption that $\boldsymbol\mu_{1:N} = \left\{\boldsymbol 0\right\}_{i=1,2,\ldots,N}$, where the pooled covariance of $X$ is redefined as 
    
    \begin{equation}
        \boldsymbol\Sigma_\ast = \sum_{i=1}^N w_i \boldsymbol\Sigma_i.
        \label{eq:pooled-cov-redef}
    \end{equation}

	The expression for $\Pi_1^\gamma\left(X\right) \geq \Pi_1^\alpha\left(X\right)$ is
	
	\begin{equation}
	    (2 \pi  e)^{\frac{n}{2}} \Det{\boldsymbol\Sigma_\ast}^\frac{1}{2}  \geq \exp\left\{\frac{1}{2}\left(n +  \sum_{i=1}^N w_i \log \left|2 \pi \boldsymbol\Sigma_i\right|\right)\right\}, 
	\end{equation}
	
    \noindent which after simplification, 
    
    \begin{equation}
        \Det{\boldsymbol\Sigma_\ast}^\frac{1}{2} \geq \prod_{i=1}^N  \Det{\boldsymbol\Sigma_i}^\frac{w_i}{2},
    \end{equation}
    
    \noindent can be appreciated to satisfy Lemma \ref{lem:power-mean-inequality}.
\end{proof}

Although Theorem \ref{thm:gmem-alphabound} highlights the reliability and flexibility of using elasticity $q=1$, we must emphasize that $q=1$ may not be the \textit{only} condition under which $\Pi_q^\gamma\left(X\right) \geq \Pi_q^\alpha\left(X\right)$, as suggested by Example \ref{ex:parametric-renyi-decomposition}. Indeed, Example \ref{ex:parametric-renyi-decomposition} suggests that the integrity of this bound on $\beta$-heterogeneity at elasticity values $q\neq1$ may depend in various ways on the unique combination of component-wise parameters in a parametric Gaussian mixture.

\begin{Example}[Decomposition of R\'enyi Heterogeneity in a Parametric Gaussian Mixture]\label{ex:parametric-renyi-decomposition} Consider a parametric Gaussian mixture $X$ with four components defined on $\mathbb R$ (for instance, Figure \ref{fig:mixturegrid}A). The components' respective standard deviations are $\boldsymbol\sigma = \left(0.5, 0.8, 1.1, 1.6\right)$. We vary the column vector of mixture component weights $\mathbf w = \left(w_i\right)_{i=1,\ldots,4}$ according to the following function,

\begin{equation}
    \mathbf w(a) = \left\{ \begin{array}{ll}
    \left(1,0,0,0\right)^\top & a = 0 \\
    \left(0.25,0.25,0.25,0.25\right)^\top & a = 1 \\
    \left(0,0,0,1\right)^\top & a = \infty \\
    \left(\frac{a^\frac{1}{3}-1}{a^\frac{4}{3} - 1} a^\frac{i-1}{3} \right)_{i=1,\ldots,4} & a \notin \{0,1,\infty\}
    \end{array}
    \right.
    \label{eq:pskew}
\end{equation}

\noindent which ``skews'' the distribution of weights over components in $X$ according to the value of a skew parameter $a \geq 0$ (shown in Figure \ref{fig:mixturegrid}B. As the parameter $a$ decreases further below 1, components $X_1$ and $X_2$ (which have the narrowest distributions) become preferentially weighted. Conversely, as $a$ increases above 1, components $X_3$ and $X_4$ are preferentially weighted.  At $a = 1$, all components are equally weighted (depicted as the dashed black lines in Figure \ref{fig:mixturegrid}B-F).

\begin{figure}[H]
	\centering
	\includegraphics[width=0.9\textwidth]{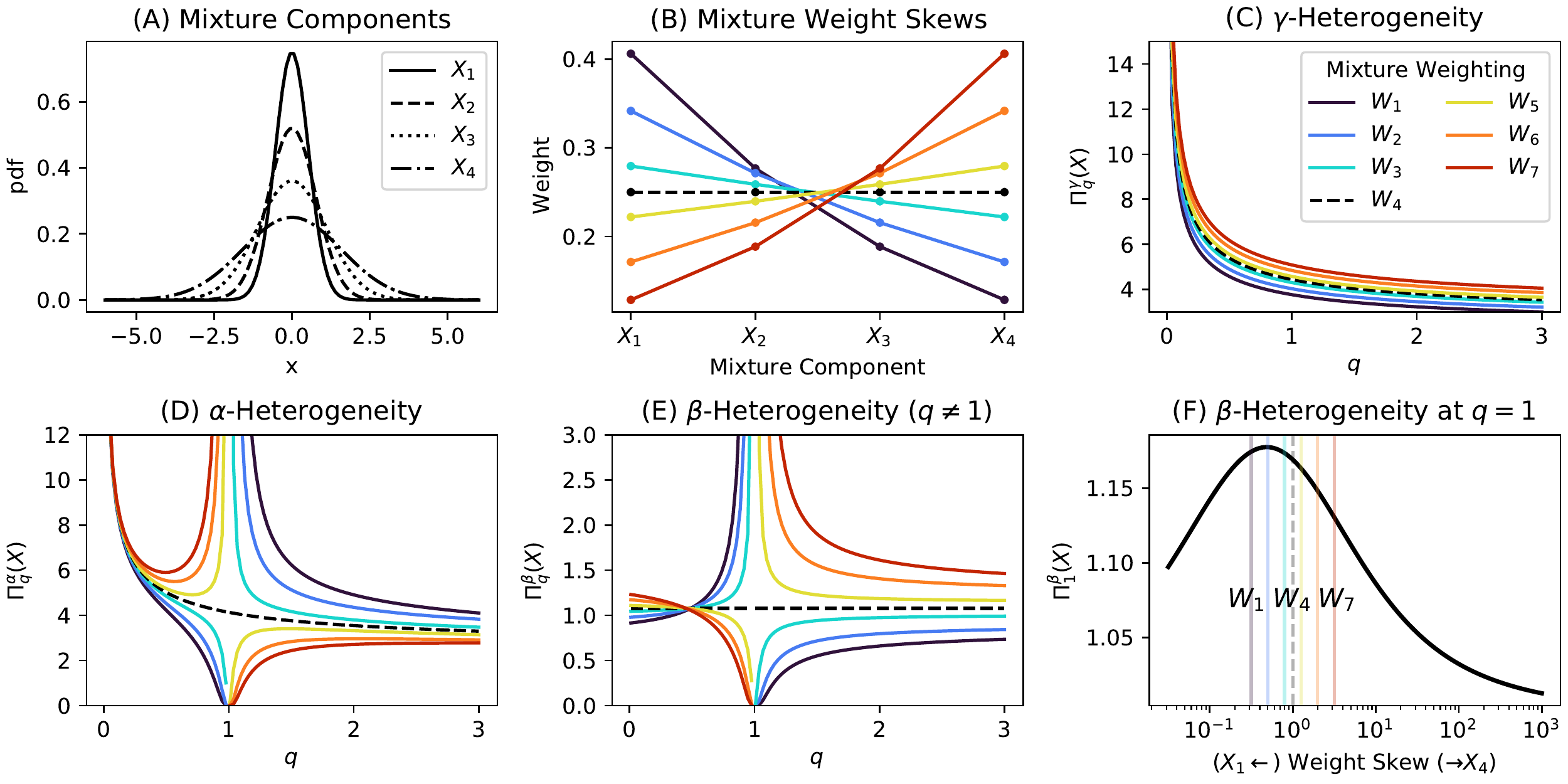}
	\caption{Graphical counterexample showing that $\alpha$-heterogeneity is not always a lower bound on $\gamma$-heterogeneity when $q\neq 1$ for a parametric Gaussian mixture. \textbf{Panel A}: Four univariate Gaussian components used in the mixture distribution evaluated. \textbf{Panel B}: Mixture component weights. Each colored line (see bottom right of Figure for legend) represents a different distribution of weights on the mixture components, such that in some settings, the most narrow components are weighted highest, and vice versa. \textbf{Panel C}: $\gamma$-heterogeneity as computed by pooling the mixture components from Panel A according to Equation \ref{eq:gaussian-global}, for each weighting scheme at $q\neq1$. \textbf{Panel D}: The $\alpha$-heterogeneity for each weighting scheme at $q\neq1$. \textbf{Panel E}: The $\beta$-heterogeneity across each weighting scheme at $q\neq1$. \textbf{Panel F}: The $\beta$-heterogeneity across various weighting schemes (plotted on the x-axis in log scale) at $q=1$. The vertical coloured lines correspond to the values of $\Pi_1^\beta\left(X\right)$ across the weighting schemes $W_{1:7}$ shown in the legend of Panel C.}
	\label{fig:mixturegrid}
\end{figure}   

Figures \ref{fig:mixturegrid}C-E plot the $\gamma$-, $\alpha$-, and $\beta$-heterogeneity for the parametric Gaussian mixture at $q \neq 1$, respectively, while Figure \ref{fig:mixturegrid}F computes the $\beta$-heterogeneity at $q=1$ for variously skewed weight distributions. When the skew parameter results in a distribution of weights whose ranking of components agrees with the rank order of component distribution widths ($\boldsymbol\sigma$), then $\beta$-heterogeneity appears to exceed 1 for $q > 1$. However, when the component weights and distribution widths are anti-correlated (in terms of rank order), then we observe values of $\beta$-heterogeneity below 1 at values of $q > 1$, as well as for some values of $q < 1$. 
\end{Example}

%%%%%%%%%%%%%%%%%%%%%%%%%%%%%%%%%%%%%%%%%%

\section{Discussion}
\label{s:discussion}

This paper provided approaches for multiplicative decomposition of heterogeneity in continuous mixture distributions, thereby extending the earlier work on discrete space heterogeneity decomposition presented by \citet{Jost2007}. Two approaches were offered, dependent upon whether the distribution over the pooled system is defined either parametrically or non-parametrically. Our results improve the understanding of heterogeneity measurement in non-categorical systems by providing conditions under which decomposition of heterogeneity into $\alpha$ and $\beta$ components conforms to the intuitive property that $\gamma \geq \alpha$. 

If one defines the pooled mixture non-parametrically, as in a finite mixture model, heterogeneity is decomposable such that $\gamma \geq \alpha$ for all $q > 0$ (if component weights are uniform, or at $q=1$ otherwise), and $\beta$ may be interpreted as the discrete number of distinct mixture components (Sections \ref{ss:continuous-renyi-decomposition} \& \ref{s:gmm}). This has the advantage of conforming with the original discrete decomposition by \citet{Jost2007}, insofar as probability mass in the mixture is recorded only where it is observed in the data, and not elsewhere, as would be assumed under a parametric model of the pooled system. Consequently, one achieves a more precise estimate of the size of the pooled system's base of support. The primary limitation arises from the need to numerically integrate the $\gamma$-heterogeneity, which can become prohibitively expensive in higher dimensions. Future work should investigate the error bounds on numerically integrated $\gamma$.

A more computationally efficient approach for decomposition of continuous R\'enyi heterogeneity is to assume that the pooled mixture has an overall parametric distribution. A common application for which this assumption is generally made is in mixed-effects meta-analysis \cite{DerSimonian1986}. An important departure from the non-parametric pooling approach of finite mixture models is that non-trivial probability mass may now be assigned to regions not covered by any of the constituent component distributions. From another perspective, one may appreciate that the non-parametric approach to pooling is insensitive to the distance between component distributions, and rather only measures the effective volume of event space to which component distributions assign probability. Conversely, assumption of the parametric distribution over mixture (in the case of Section \ref{s:hier-gaussian-random-effects}, a Gaussian) incorporates the distance between the component distributions into the calculation of $\gamma$-heterogeneity. This would be appropriate in scenarios where one assumes that the observed components undersamples the true distribution on the pooled system. For example, in the case of mixed-effects meta-analysis, the available research studies for inclusion may differ significantly in terms of their means, but one might assume that there is a significant probability of a new study yielding an effect somewhere in between. Specifying a parametric distribution over the pooled system would capture this assumption.  

One limitation of the present study is the use of a Gaussian model for the pooled system distribution. This was chosen on account of (A) its prevalence in the scientific literature and (B) analytical tractability. Future work should expand these results to other distributions. Notwithstanding, we have demonstrated the decomposition of $\gamma$ R\'enyi heterogeneity into its $\alpha$ and $\beta$ components for continuous systems. There are (broadly) two approaches, based on whether parametric assumptions are made about the pooled system distribution. Under these assumptions applied to Gaussian mixture distributions, we provided conditions under which the criterion that $\gamma \geq \alpha$ is satisfied. Future studies should evaluate this method as an alternative approach for the measurement of meta-analytic heterogeneity, and expand these results to other parametric distributions over the pooled system.

%%%%%%%%%%%%%%%%%%%%%%%%%%%%%%%%%%%%%%%%%%
%\section{Conclusions}

%This section is not mandatory, but can be added to the manuscript if the discussion is unusually long or complex.

%%%%%%%%%%%%%%%%%%%%%%%%%%%%%%%%%%%%%%%%%%
%\section{Patents}
%This section is not mandatory, but may be added if there are patents resulting from the work reported in this manuscript.

%%%%%%%%%%%%%%%%%%%%%%%%%%%%%%%%%%%%%%%%%%
%\vspace{6pt} 

%%%%%%%%%%%%%%%%%%%%%%%%%%%%%%%%%%%%%%%%%%
%% optional
%\supplementary{The following are available online at \linksupplementary{s1}, Figure S1: title, Table S1: title, Video S1: title.}


%%%%%%%%%%%%%%%%%%%%%%%%%%%%%%%%%%%%%%%%%%
\authorcontributions{Conceptualization, A.N.; methodology, A.N.;  validation, A.N.; formal analysis, A.N.; investigation, A.N.; writing--original draft preparation, A.N.; writing--review and editing, M.A. and T.T.; visualization, A.N.; supervision, M.A. and T.T.}

%%%%%%%%%%%%%%%%%%%%%%%%%%%%%%%%%%%%%%%%%%
\funding{This research received no external funding.}

%%%%%%%%%%%%%%%%%%%%%%%%%%%%%%%%%%%%%%%%%%
%\acknowledgments{In this section you can acknowledge any support given which is not covered by the author contribution or funding sections. This may include administrative and technical support, or donations in kind (e.g., materials used for experiments).}

%%%%%%%%%%%%%%%%%%%%%%%%%%%%%%%%%%%%%%%%%%
\conflictsofinterest{The authors declare no conflict of interest.} 

%%%%%%%%%%%%%%%%%%%%%%%%%%%%%%%%%%%%%%%%%%
%% optional

%%%%%%%%%%%%%%%%%%%%%%%%%%%%%%%%%%%%%%%%%%
%% optional
\appendixtitles{yes} %Leave argument "no" if all appendix headings stay EMPTY (then no dot is printed after "Appendix A"). If the appendix sections contain a heading then change the argument to "yes".
\appendix
\section{Proofs}\label{app:proofs}
\unskip

\begin{proof}[Proof of Theorem \ref{thm:decomposition-nonparametric-continuous}] Following \citet{Jost2007} (proof 2), in the limit $q\to 1$, one obtains the following inequality
    
\begin{equation}
	 -\sum_{i=1}^{N} w_i \int_{\mathcal X} f_i(\mathbf x) \log f_i(\mathbf x) \diff \mathbf x \leq -\int_{\mathcal X} \fbar(\mathbf x) \log \fbar(\mathbf x) \diff \mathbf x,
\end{equation} 

\noindent whereas when $w_i = w_j$ for all $(i,j)\in \{1,2,\ldots,N\}$, for $q > 1$ we have
    
\begin{equation}
 \frac{1}{N} \sum_{i=1}^{N} \int_{\mathcal X} f_i^q(\mathbf x) \diff \mathbf x \geq \int_{\mathcal X} \left(\frac{1}{N} \sum_{i=1}^N f_i(\mathbf x)\right)^q \diff \mathbf x.
\end{equation}

\noindent and for $q<0$ we have 

\begin{equation}
 \frac{1}{N} \sum_{i=1}^{N} \int_{\mathcal X} f_i^q(\mathbf x) \diff \mathbf x \leq \int_{\mathcal X} \left(\frac{1}{N} \sum_{i=1}^N f_i(\mathbf x)\right)^q \diff \mathbf x,
\end{equation}

\noindent all of which hold by Jensen's inequality.
\end{proof}

\begin{proof}[Proof of Proposition \ref{prop:gaussian-renyi-het}.] We must solve the following integral: 

\begin{equation}
\Pi_q\left(X\right) = \left[
        \left(2 \pi\right)^{-\frac{qn}{2}} 
        \left|\boldsymbol\Sigma\right|^{-\frac{q}{2}}
        \int_{\mathbb R^n}  e^{-\frac{q}{2}\left(\mathbf x - \boldsymbol\mu\right)^\top \boldsymbol\Sigma^{-1} \left(\mathbf x - \boldsymbol\mu\right)} \diff \mathbf x
    \right]^\frac{1}{1-q}
\label{eq:gaussian-qintegral}
\end{equation}

The eigendecomposition of the inverse of the covariance matrix $\boldsymbol\Sigma^{-1}$ into an orthonormal matrix of eigenvectors $\mathbf U$ and an $n \times n$ diagonal matrix of eigenvalues $\boldsymbol\Lambda = \left(\delta_{ij} \lambda_i \right)_{i=1,2,\ldots,n}^{j=1,2,\ldots,n}$, where $\delta_{ij}$ is Kronkecker's delta, facilitates the substitution $\mathbf y = \mathbf U^{-1}(\mathbf x - \boldsymbol\mu)$ required for Gaussian integration, by which we obtain the following solution for $q \notin \{0,1,\infty\}$:

\begin{equation}
    \Pi_q\left(X\right) = q^{\frac{n}{2(q-1)}} (2 \pi )^{\frac{n}{2}} \Det{\boldsymbol\Sigma}^\frac{1}{2}.
    \label{eq:gaussian-qintegral-solution}
\end{equation}

L'H\^opital's rule facilitates computation of the limit as $q\to1$:

\begin{equation}
\begin{split}
    \lim_{q\to 1} \log \Pi_q\left(X\right) &= \lim_{q\to1} \left(\frac{n}{2(q-1)}\log q\right) + \frac{n}{2}\log (2\pi) + \frac{1}{2}\log \left|\boldsymbol\Sigma\right| \\
    &= \frac{n}{2} + \frac{n}{2}\log (2\pi) + \frac{1}{2}\log \left|\boldsymbol\Sigma\right|, \\
\end{split}
\end{equation}

\noindent giving the \textit{perplexity},

\begin{equation}
    \Pi_1\left(X\right) = \left(2 \pi e\right)^\frac{n}{2} \Det{\boldsymbol\Sigma}^\frac{1}{2}.
    \label{eq:renymvn-q1}
\end{equation}

\noindent By the same procedure, we can compute the limit as $q\to\infty$, 

\begin{equation}
    \Pi_\infty\left(X\right) = \left(2 \pi \right)^\frac{n}{2} \Det{\boldsymbol\Sigma}^\frac{1}{2},
    \label{eq:renymvn-qinf}
\end{equation}

\noindent as well as show that $\Pi_0\left(X\right)$ is undefined.
\end{proof}

\begin{proof}[Proof of Lemma \ref{lem:gamma-dispersion}.]
    For all $q > 0$, proving $\Pi_q^\gamma\left(X^\prime\right) \geq \Pi_q^\gamma\left(X\right)$ amounts to proving $\Det{\boldsymbol\Sigma_\ast^\prime}^\frac{1}{2} \geq \Det{\boldsymbol\Sigma_\ast}^\frac{1}{2}$. To this end, we have 
    
    \begin{align}
        \boldsymbol\Sigma_\ast^\prime 
            &= \sum_{i=1}^N w_i \boldsymbol\Sigma_i 
            + \left[ \sum_{i=1}^N w_i \left(\sqrt{c}\boldsymbol\mu_i\right)\left(\sqrt{c}\boldsymbol\mu_i\right)^\top - \left(\sum_{i=1}^N w_i\sqrt{c}\boldsymbol\mu_i\right)\left(\sum_{i=1}^N w_i \sqrt{c}\boldsymbol\mu_i\right)^\top \right] \\
            &= \sum_{i=1}^N w_i \boldsymbol\Sigma_i + c \left(\sum_{i=1}^N w_i \boldsymbol\mu_i \boldsymbol\mu_i^\top - \boldsymbol\mu_\ast \boldsymbol\mu_\ast^\top \right) \\
            &= \hat{\boldsymbol\Sigma} + c \mathbf C[\boldsymbol\mu]
            \label{eq:covprime-equivalence}
    \end{align}
    
    \noindent and 
    
    \begin{equation}
    \boldsymbol\Sigma_\ast = \hat{\boldsymbol\Sigma} + \mathbf C[\boldsymbol\mu],
    \end{equation}

    \noindent where we denoted $\hat{\boldsymbol\Sigma} = \sum_{i=1}^N w_i \boldsymbol\Sigma_i$ and $\mathbf C[\boldsymbol\mu] = \sum_{i=1}^N w_i \boldsymbol\mu_i \boldsymbol\mu_i^\top - \boldsymbol\mu_\ast \boldsymbol\mu_\ast^\top$ for notational parsimony. Clearly, when $c=1$, we have $\boldsymbol\Sigma_\ast^\prime = \boldsymbol\Sigma_\ast$.
    
    By the Minkowski determinant inequality, we have that 
    
    \begin{align}
    \Det{\boldsymbol\Sigma_\ast^\prime}^\frac{1}{2} &\geq \Det{\hat{\boldsymbol\Sigma}}^\frac{1}{2} + c^\frac{n}{2}\Det{\mathbf C[\boldsymbol\mu]}^\frac{1}{2} \\
    \Det{\boldsymbol\Sigma_\ast}^\frac{1}{2} &\geq \Det{\hat{\boldsymbol\Sigma}}^\frac{1}{2} + \Det{\mathbf C[\boldsymbol\mu]}^\frac{1}{2},
    \end{align}
    
    \noindent which, since $c\geq 1$, implies the first line is greater than or equal to the second. Subtracting the second line from the first and simplifying yields 
    
    \begin{equation}
        \frac{\Det{\boldsymbol\Sigma_\ast^\prime}^\frac{1}{2} - \Det{\boldsymbol\Sigma_\ast}^\frac{1}{2}}{\Det{\mathbf C[\boldsymbol\mu]}^\frac{1}{2}} \geq c^\frac{n}{2} - 1
        \label{eq:lem1-finaleq}
    \end{equation}
    
    \noindent At $c=1$ Equation \ref{eq:lem1-finaleq} reduces to an equality, and since $c\geq 1$ and $n \geq 1$, Equation \ref{eq:lem1-finaleq} establishes that $\Det{\boldsymbol\Sigma_\ast^\prime}^\frac{1}{2} \geq \Det{\boldsymbol\Sigma_\ast}^\frac{1}{2}$.
\end{proof}

\begin{proof}[Proof of Lemma \ref{lem:power-mean-inequality}.]
    Since $\boldsymbol\Sigma_{1:N}$ are positive semidefinite matrices, then for all $\mathbf x \in \mathbb R^n$, we have that $- \frac{1}{2} \mathbf x^\top \left(w_i \boldsymbol\Sigma_i \right) \mathbf x \leq 0$, and thus $- \frac{1}{2} \mathbf x^\top \left(\sum_{i=1}^N w_i \boldsymbol\Sigma_i \right) \mathbf x \leq 0$. By exponentiating the quadratic term, we have 
    
    \begin{equation}
        e^{- \frac{1}{2} \mathbf x^\top \left(\sum_{i=1}^N w_i \boldsymbol\Sigma_i \right) \mathbf x } = \prod_{i=1}^N \left(e^{- \frac{1}{2} \mathbf x^\top \boldsymbol\Sigma_i \mathbf x}\right)^{w_i}.
        \label{eq:exponentiated}
    \end{equation}
    
    We obtain the following expressions by applying Gaussian integration to the left hand side,
    
    \begin{equation}
        \int_{\mathbb R^n} e^{- \frac{1}{2} \mathbf x^\top \left(\sum_{i=1}^N w_i \boldsymbol\Sigma_i \right) \mathbf x } \diff \mathbf x = \left(2 \pi\right)^\frac{n}{2} \Det{\left(\sum_{i=1}^N w_i \boldsymbol\Sigma_i \right)}^{-\frac{1}{2}},
        \label{eq:gauss-int-lhs}
    \end{equation}
    
    \noindent as well as to a bound on the right hand side obtained by H\"older's inequality,  
    
    \begin{align}
         \int_{\mathbb R^n} \prod_{i=1}^N \left(e^{-\frac{1}{2}\mathbf x^\top \boldsymbol\Sigma_i \mathbf x}\right)^{w_i} \diff \mathbf x 
            &\leq
            \prod_{i=1}^N \left(\int_{\mathbb R^n} e^{-\frac{1}{2}\mathbf x^\top \boldsymbol\Sigma_i \mathbf x} \diff \mathbf x\right)^{w_i} \\
            &= \left(2\pi\right)^\frac{n}{2} \left(\prod_{i=1}^N \Det{\boldsymbol\Sigma_i}^{-\frac{w_i}{2}}\right).
        \label{eq:gauss-int-rhs}
    \end{align}
    
    \noindent Substituting Equations \ref{eq:gauss-int-lhs} and \ref{eq:gauss-int-rhs} into Equation \ref{eq:exponentiated} and simplifying terms yields
    
    \begin{equation}
        \Det{\sum_{i=1}^N w_i \boldsymbol\Sigma_i}^\frac{1}{2} \geq \prod_{i=1}^N \Det{\boldsymbol\Sigma_i}^\frac{w_i}{2}.
    \end{equation}
\end{proof}


%=====================================
% References, variant B: external bibliography
%=====================================
\reftitle{References}
\externalbibliography{yes}
\bibliography{bibliography}

%%%%%%%%%%%%%%%%%%%%%%%%%%%%%%%%%%%%%%%%%%
%% optional
%\sampleavailability{Samples of the compounds ...... are available from the authors.}

%% for journal Sci
%\reviewreports{\\
%Reviewer 1 comments and authors’ response\\
%Reviewer 2 comments and authors’ response\\
%Reviewer 3 comments and authors’ response
%}

%%%%%%%%%%%%%%%%%%%%%%%%%%%%%%%%%%%%%%%%%%
\end{document}